\documentclass[10pt,english,amsmath,amssymb,prl,showkeys,superscriptaddress,twocolumn,showpacs,floatfix]{revtex4-2}
\usepackage{graphicx}
\usepackage{tabularx}
\usepackage{epstopdf}
\usepackage{dcolumn}
\usepackage{chemformula}
\usepackage[colorlinks=true,linkcolor=blue ,citecolor=blue,urlcolor=blue]{hyperref}
\usepackage{multirow}
\usepackage[usenames,dvipsnames]{xcolor}
\usepackage{soul}
\usepackage{braket}
\usepackage{bm}
\usepackage{nicefrac}
\usepackage{siunitx}
\usepackage{color,transparent}
\usepackage[utf8]{inputenc}
\usepackage{upgreek}

\usepackage{amsmath}   
\usepackage{mathtools} 
\usepackage{amssymb}   
\usepackage{mathrsfs}  
\usepackage{textcomp}  
\usepackage{accents}   
\usepackage{bm}        
\usepackage{soul}
\usepackage{relsize}
\usepackage{graphicx}  
\usepackage{epstopdf}  
\usepackage{subfigure} 
\usepackage{makecell}
\usepackage{hhline}
\usepackage{cellspace}
\usepackage{array}     
\usepackage{tabularx}  
\usepackage{multirow}  
\usepackage{booktabs}  
\usepackage{dcolumn}   
\usepackage{gensymb}   
\usepackage{setspace}
\usepackage{scrextend}
\usepackage{float}

\usepackage{chemformula}

\begin{document}
\setchemformula{kroeger-vink}
\setlength{\heavyrulewidth}{0.08em}
\setlength{\lightrulewidth}{0.05em}
\setlength{\cmidrulewidth}{0.03em}
\setlength{\belowrulesep}{0.65ex}
\setlength{\belowbottomsep}{0.00pt}
\setlength{\aboverulesep}{0.40ex}
\setlength{\abovetopsep}{0.00pt}
\setlength{\cmidrulesep}{\doublerulesep}
\setlength{\cmidrulekern}{0.50em}
\setlength{\defaultaddspace}{0.50em}
\setlength{\tabcolsep}{4pt}

\title{Why hole polaron formation on oxygen is limiting the Fermi level in  Fe acceptor doped BaTiO$_{3}$ under oxidizing conditions}
\author{Mohammad Amirabbasi}
\email{amirabbasi@mm.tu-darmstadt.de}
\address{Institute of Materials Science, Materials Modelling, Technical University of Darmstadt, Otto-Berndt-Str. 3, Darmstadt 64283, Germany}

\author{Emre Erdem}
\address{Center of Excellence for Functional Surfaces and Interfaces for Nano-Diagnostics (EFSUN), Sabancı University, Tuzla, 34956, Istanbul, Turkey}
\address{Faculty of Engineering and Natural Sciences, Sabanci University, Tuzla, 34956 Istanbul, Turkey}
\author{Denis Sudarikov}
\address{Institute of Materials Science, Electronic Structure of Materials, Technical University of Darmstadt, Otto-Berndt-Str 3, Darmstadt 64287, Germany}
\author{Jochen Rohrer}
\address{Institute of Materials Science, Materials Modelling, Technical University of Darmstadt, Otto-Berndt-Str. 3, Darmstadt 64283, Germany}
\author{Andreas Klein}
\address{Institute of Materials Science, Electronic Structure of Materials, Technical University of Darmstadt, Otto-Berndt-Str 3, Darmstadt 64287, Germany}
\author{Karsten Albe}
\email{albe@mm.tu-darmstadt.de}
\address{Institute of Materials Science, Materials Modelling, Technical University of Darmstadt, Otto-Berndt-Str. 3, Darmstadt 64283, Germany}

\begin{abstract}
Oxidizing Fe-doped BaTiO$_3$ is commonly expected to convert substitutional
Fe$^{3+}$ acceptors into formal Fe$^{4+}$ centers. Yet, the experimentally
accessible picture based on electron-paramagnetic resonance (EPR) is dominated by Fe$^{3+}$-related signatures, while Fe$^{4+}$
is not a straightforward observable. Here we show that this apparent
discrepancy reflects the preferred location of the oxidizing hole: not on Fe, but
on oxygen. Using density-functional theory with with occupation-matrix control and a
piecewise-linearity-based Hubbard correction (DFT+$U$) for O-2$p$ states, we find that an
oxygen-centered hole polaron is forming a
Fe$^{3+}$--O$^{-}$ complex that is lower in energy than the formal Fe$^{4+}$
configuration. Our  results identify ligand-hole formation as a favorable
charge-compensation mechanism in oxidized Fe-doped BaTiO$_3$ and provide an
explanation for the predominance of Fe$^{3+}$-based centers in
spectroscopy. More broadly, they show how oxygen polarons can limit Fermi-level
shifts and control the electronic response of acceptor-doped ferroelectric
perovskites.
\end{abstract}

\maketitle
\textit{Introduction--} 
Chemical doping is a central strategy for tailoring the functional properties of BaTiO$_3$, including its dielectric response, insulation resistance, and defect transport behavior~\cite{Iqbal2025,Acosta_2017,Adediji2023}. In acceptor-doped BaTiO$_3$, the technologically relevant properties are governed not only by the nominal dopant concentration, but also by the charge-compensation mechanism that determines the position of the Fermi level. In multilayer ceramic capacitors, for example, transition-metal acceptors such as Fe are widely used to control electronic conductivity and degradation under bias~\cite{Yang2004,suzuki2019energy}. A substitutional Fe impurity on the Ti site is commonly described as an acceptor,
Fe$_{\mathrm{Ti}}^{'}$, when Fe is trivalent, whereas oxidation under sufficiently low Fermi-level conditions would formally lead to the isovalent Fe$_{\mathrm{Ti}}^{\times}$ state, corresponding to Fe$^{4+}$ on the Ti$^{4+}$ sublattice.

This simple ionic picture, however, is difficult to assess directly by spectroscopy. Electron-paramagnetic resonance (EPR), for example, is highly sensitive to high-spin
Fe$^{3+}$ centers and their local defect environments, and previous studies of
Fe-doped BaTiO$_3$ and related titanate perovskites consistently identify
Fe$^{3+}$-based centers, often associated with oxygen-related defects~\cite{schwartz1993electron,possenriede1992paramagnetic,laguta2005electron,theerthan2012magnetic}.
By contrast, Fe$^{4+}$ cannot be used as a straightforward EPR benchmark: depending on its spin state, covalency, and relaxation dynamics, it may be EPR silent or produce a signal that is too broad or weak to assign unambiguously. Thus, the lack of a
resolved Fe$^{4+}$ signal does not prove its absence. This motivates the central question addressed here: under oxidizing conditions, is the compensating hole accommodated by oxidation of Fe, or does it localize on a neighboring oxygen ligand to form a Fe$^{3+}$--O$^{-}$ bound-polaron complex?

\begin{figure}[!htbp]
  \centering

  \subfigure[Fe$^{4+}$--O$^{2-}$]{%
    \includegraphics[width=0.40\columnwidth]{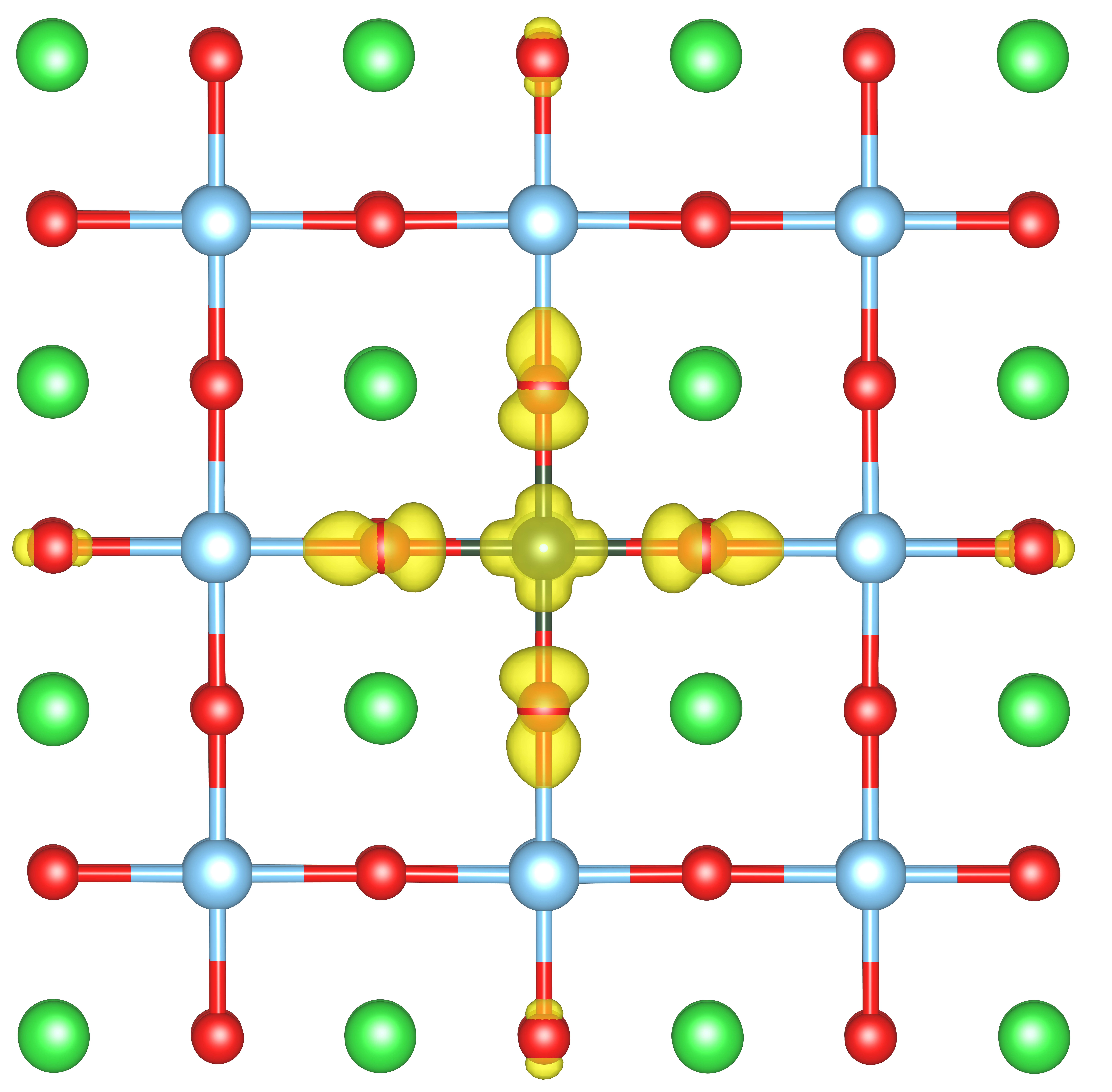}%
    \label{fig:Fig-a}%
  }
  \hfill
  \includegraphics[width=0.15\columnwidth]{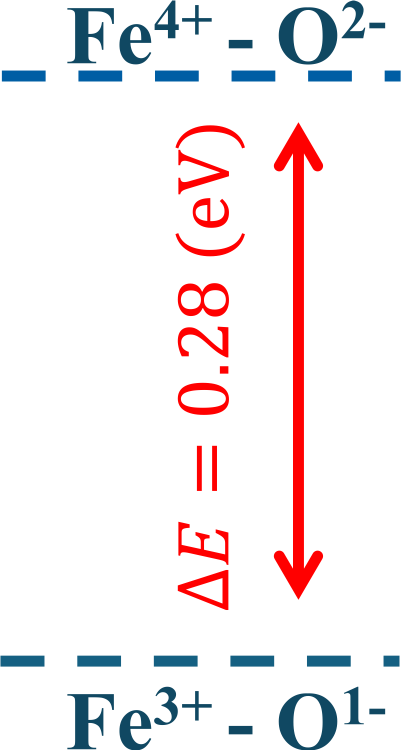}
  \hfill
  \subfigure[Fe$^{3+}$--O$^{-}$]{%
    \includegraphics[width=0.40\columnwidth]{Figure/Fig-c.pdf}%
    \label{fig:Fig-b}%
  }
  \caption{(Color online) {Two charge-compensation configurations for substitutional Fe on the Ti site in \ch{BaTiO3}.
(a) \ch{Fe^{4+}} with a delocalized excess hole on the Fe--O network (yellow isosurface).
(b) \ch{Fe^{3+}} associated with an O-centred hole polaron on a nearest-neighbour oxygen (\ch{Fe^{3+}}--\ch{O^-} bound polaron).
Atom colours: Ba, green; Ti, blue; O, red; and Fe, black.
The \ch{Fe^{3+}}--\ch{O^-} complex is energetically more favourable than the \ch{Fe^{4+}} configuration by $\Delta E = 0.28$~eV.}}
\label{Fig-graphical}
\end{figure}

Oxygen-centered holes are well known in oxide materials, where strong electron--lattice coupling can localize a hole on an oxygen ion and form a small
hole polaron~\cite{Schirmer_2006,Schirmer_2011,Franchini2021}.   In BaTiO$_3$, EPR studies of alkali acceptors have shown that negatively charged A-site
dopants can be compensated by holes localized on neighboring oxygen ligands, forming bound O$^{-}$ polarons~\cite{Varnhorst1996,Schirmer_2011}. These results
suggest that the relevant oxidized state of acceptor-doped BaTiO$_3$ may be a charge-transfer configuration involving an O-centered ligand hole rather than a
pure high-valent transition-metal ion. For Fe doping, this possibility is particularly important because Fe--O covalency can blur the distinction between
formal Fe$^{4+}$ and Fe$^{3+}\underline{L}$ configurations, where $\underline{L}$ denotes a ligand hole.

The formation of such bound hole polarons is also directly relevant to Fermi-level engineering \cite{Klein2023Fermi}. A localized polaronic state introduces an energetically accessible charge-compensation channel that can limit further Fermi-level shifts by stabilizing electronic charge on a specific lattice site or defect complex. Therefore, identifying whether an oxidizing hole resides on Fe or on a neighboring oxygen ligand is essential for understanding both the defect chemistry and the electronic conductivity of Fe-doped BaTiO$_3$.


In this work, we use density-functional theory with Hubbard corrections
(DFT+$U$) and occupation-matrix control to compare two limiting 
charge-compensation mechanisms for substitutional Fe on the Ti site in
rhombohedral BaTiO$_3$: a formal Fe$^{4+}$ configuration and a Fe$^{3+}$ center
bound to an oxygen-centered hole polaron, Fe$^{3+}$--O$^{-}$. The Hubbard
correction applied to the O-2$p$ states is chosen by enforcing the
piecewise-linearity condition for the localized hole-polaron level. We first
establish the stability of an isolated O-centered hole polaron in BaTiO$_3$ and
then determine whether Fe substitution favors oxidation of the Fe center or
localization of the compensating hole on a neighboring oxygen ligand. Our results
show that the Fe$^{3+}$--O$^{-}$ bound-polaron complex is energetically favored
over the formal Fe$^{4+}$ state. This provides a microscopic explanation for the
dominance of Fe$^{3+}$-based EPR signatures in Fe-doped BaTiO$_3$ and identifies
oxygen hole-polaron formation as a key charge-compensation mechanism under
oxidizing conditions.
All computational details, as well as information regarding how to optimize the primitive cell and calculate dielectric constant and trapping/binding energy of the bound polaron are provided in the supplementary material~\cite{KRESSE199615, Kresse_1999, Bloch_1994, Perdew_1996, Perdew_2008, Villars2023:sm_isp_sd_1900558, Gajdo,Wu,Baroni2001, Falletta_2020, Franchini2021, Paul2014, Osterbacka2020, allen2014occupation, Wang:01, Wu2005, Moore2024}.

\begin{figure}[!htbp]
  \centering
  \subfigure[]{
    \includegraphics[width=0.48\columnwidth]{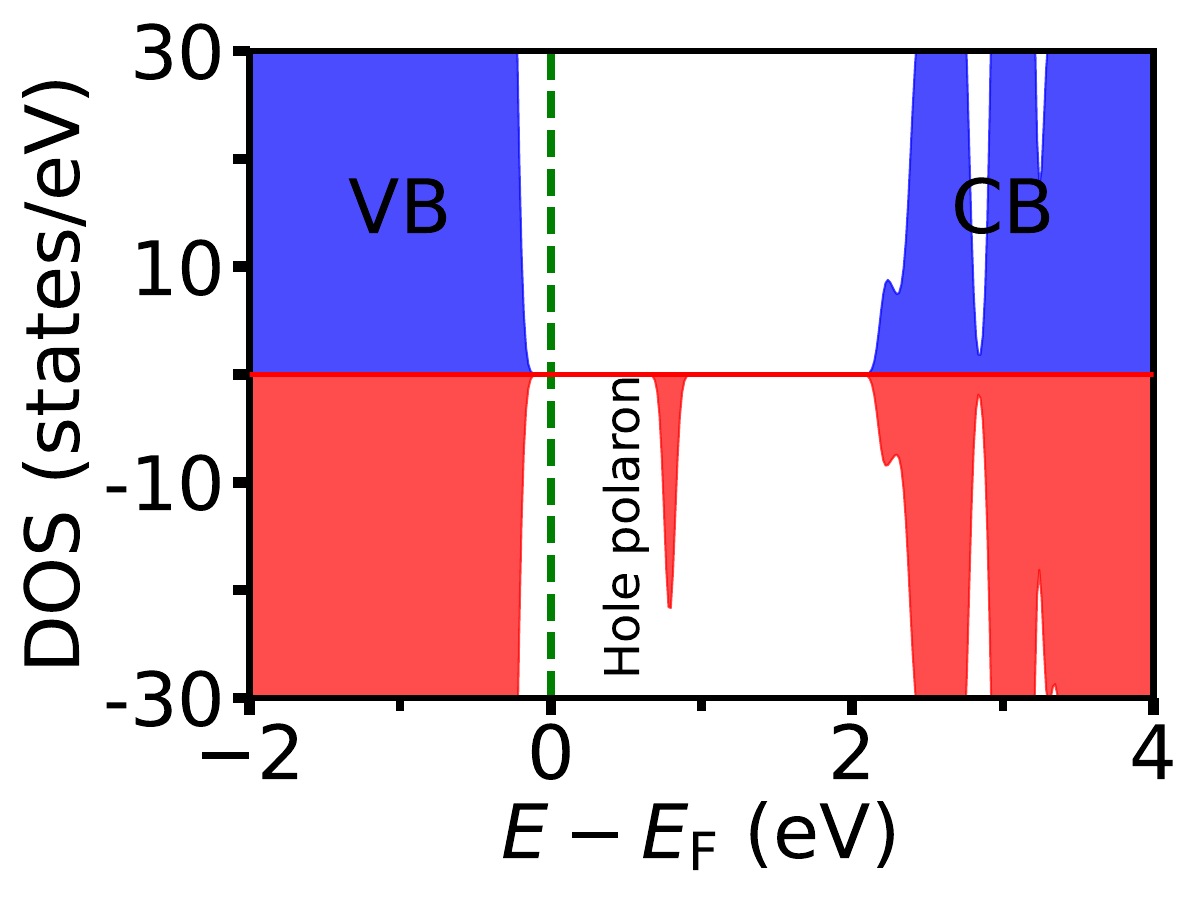}\label{fig:1-1}}
  \subfigure[]{
    \includegraphics[width=0.48\columnwidth]{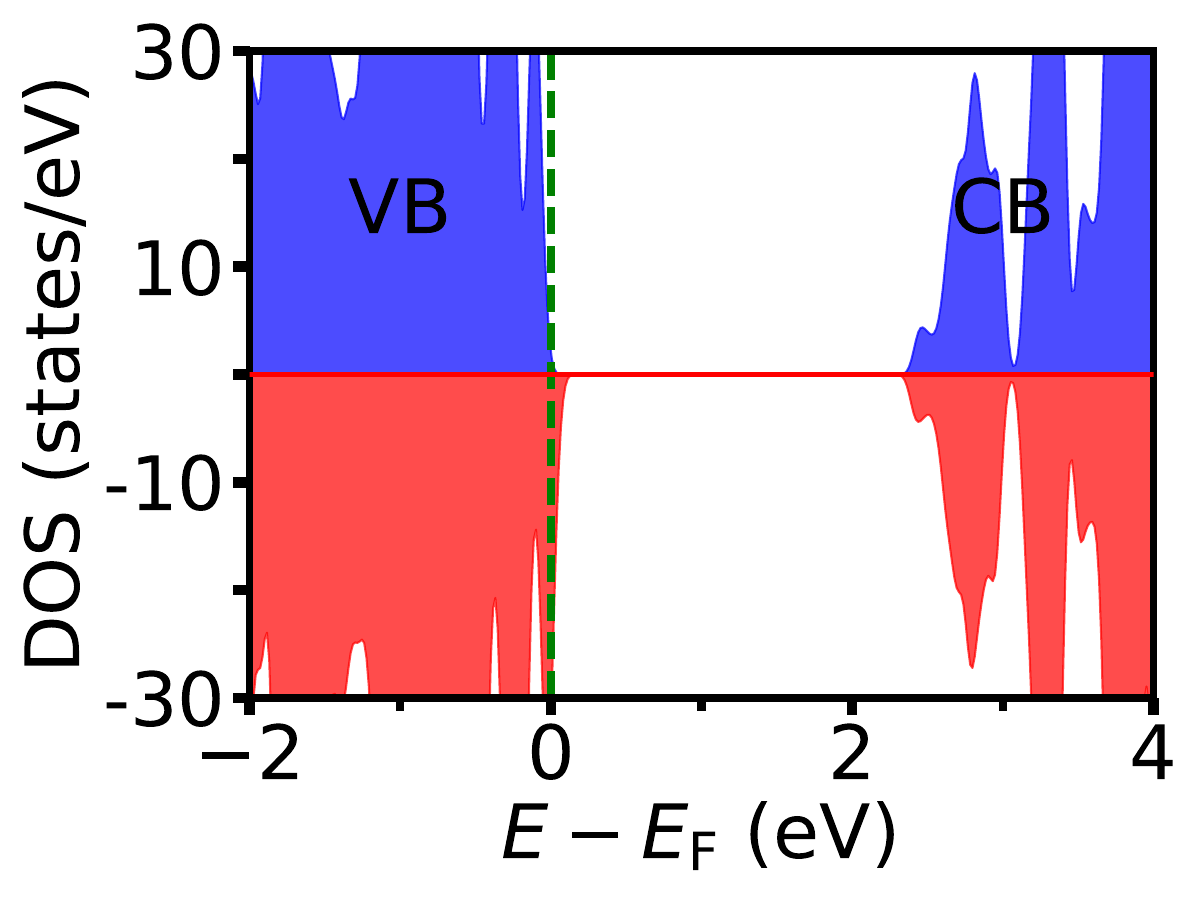}\label{fig:1-2}}
  \caption{(Color online) Density of states (DOS) for (a) a localized small hole polaron on O-2$p$ and (b) a delocalized hole.
Blue and red colors denote the spin‑up and spin‑down channels, respectively.   The Fermi energy is set to zero, and the hole-polaron level lies 1.0~(eV) with respect to the VBM.
}
  \label{fig:fig-2}
\end{figure}

\textit{Hole polaron on O-center in pristine BaTiO$_3$--}  
Before addressing Fe-doped BaTiO$_3$, we first establish whether a hole can self-trap on oxygen in the pristine host. This step is essential because a localized O-centered hole provides the reference state for the Fe$^{3+}$--O$^{-}$ bound-polaron complex discussed below.  In a conventional band picture, removal of one electron creates a delocalized hole at the valence-band maximum, which is dominated by O-2$p$ states. In an oxide with sufficiently strong electron--lattice coupling, however, the same hole may localize on a single
oxygen ligand, accompanied by a local structural distortion. The stable charge state is then not a free valence-band hole, but a small O$^{-}$ polaron \cite{alexandrov2010advances,emin2013polarons,Franchini2021,Schirmer_2006,Schirmer_2011}. 

A reliable comparison between these two solutions requires a treatment that does not artificially favor charge delocalization. Semilocal density functionals suffer from self-interaction error and therefore tend to overstabilize delocalized holes, which can suppress the symmetry breaking required for small-polaron formation \cite{Erhart2014HolePolarons,Franchini2021}. We therefore employ DFT+$U$ and apply a Hubbard correction to the O-2$p$ manifold that hosts the hole. The value of $U_{\mathrm{O}\,2p}$ is determined by enforcing the piecewise-linearity condition for the localized polaronic state, using the finite-size correction procedure of Falletta and co-workers (see Fig. S1 ) \cite{Falletta_2020,Falletta2022,Janak1978}.  This yields $U_{\mathrm{O}\,2p}=7.80$ eV, which is used throughout the subsequent
calculations. A Hubbard correction of $U_{\mathrm{Ti}\,3d}=4.5$ eV is applied to the Ti-3$d$ states \cite{Moore2024}. 
 
Starting from a rhombohedral BaTiO$_3$ supercell with one electron removed, we initialize two distinct electronic configurations: a delocalized valence-band hole and a localized hole on one oxygen site. The localized solution is obtained by imposing an initial occupation-matrix constraint on the selected O-2$p$ orbitals and allowing the lattice to relax around the hole. After convergence, the constraint is released and the system is relaxed again. The persistence of the localized solution after removing the constraint confirms that the polaronic state is a genuine local minimum rather than an artifact of the initialization procedure \cite{allen2014occupation,Osterbacka2020}.

The resulting density of states is shown in Fig. \ref{fig:fig-2}. For the localized solution, a spin-polarized O-2$p$ state appears in the band gap, approximately $1.0$ eV above the valence-band maximum. The corresponding spin density and lattice relaxation are concentrated on one oxygen ligand, identifying the state as an O-centered small hole polaron. In contrast, the delocalized solution shows no isolated in-gap O-2$p$ level; the hole remains distributed over the valence-band manifold.

The energetic stability of the self-trapped hole is quantified by the trapping energy~\cite{Osterbacka2020}:
\begin{equation}
E_{\mathrm{trap}} = E_{\mathrm{loc}}^{+1} - E_{\mathrm{pristine}}^{0}
+ \epsilon_{\mathrm{VBM}}+ E_{\mathrm{correction}}
\end{equation}
where \(E_{\mathrm{loc}}^{+1}\) is the total energy of the localized hole configuration,
\(E_{\mathrm{pristine}}^{0}\) is the total energy of the neutral
pristine configuration, \(\epsilon_{\mathrm{VBM}}\) denotes the energy
of the valence-band maximum of the pristine system, and $E_{\mathrm{corr}}$ is the finite-size correction, defined in the Supplemental Material, with a value of 0.04 eV.
We obtain 
\begin{equation}
E_{\mathrm{trap}}=-0.17~\mathrm{eV},
\end{equation}
showing that localization of the hole on oxygen is energetically favored. The negative trapping energy establishes that pristine rhombohedral BaTiO$_3$ can stabilize an O-centered hole polaron.

Finally, we test whether this conclusion depends sensitively on the Hubbard correction applied to Ti-3$d$ states. Repeating the calculations for $U_{\mathrm{Ti}\,3d}=3.5$, $4.0$, and $5.0$ eV changes the trapping energy only from $-0.19$ to $-0.16$ eV. The variation remains within $0.02$ eV of the value obtained for $U_{\mathrm{Ti}\,3d}=4.5$ eV. Thus, the predicted stability of the oxygen hole polaron is not controlled by the precise choice of the Ti Hubbard parameter. This stable O$^{-}$ polaron provides the necessary host reference for evaluating whether an oxidizing hole in Fe-doped BaTiO$_3$ resides on Fe or on a neighboring oxygen ligand.

\begin{figure}[!htbp]
  \centering
  \subfigure[]{
    \includegraphics[width=0.48\columnwidth]{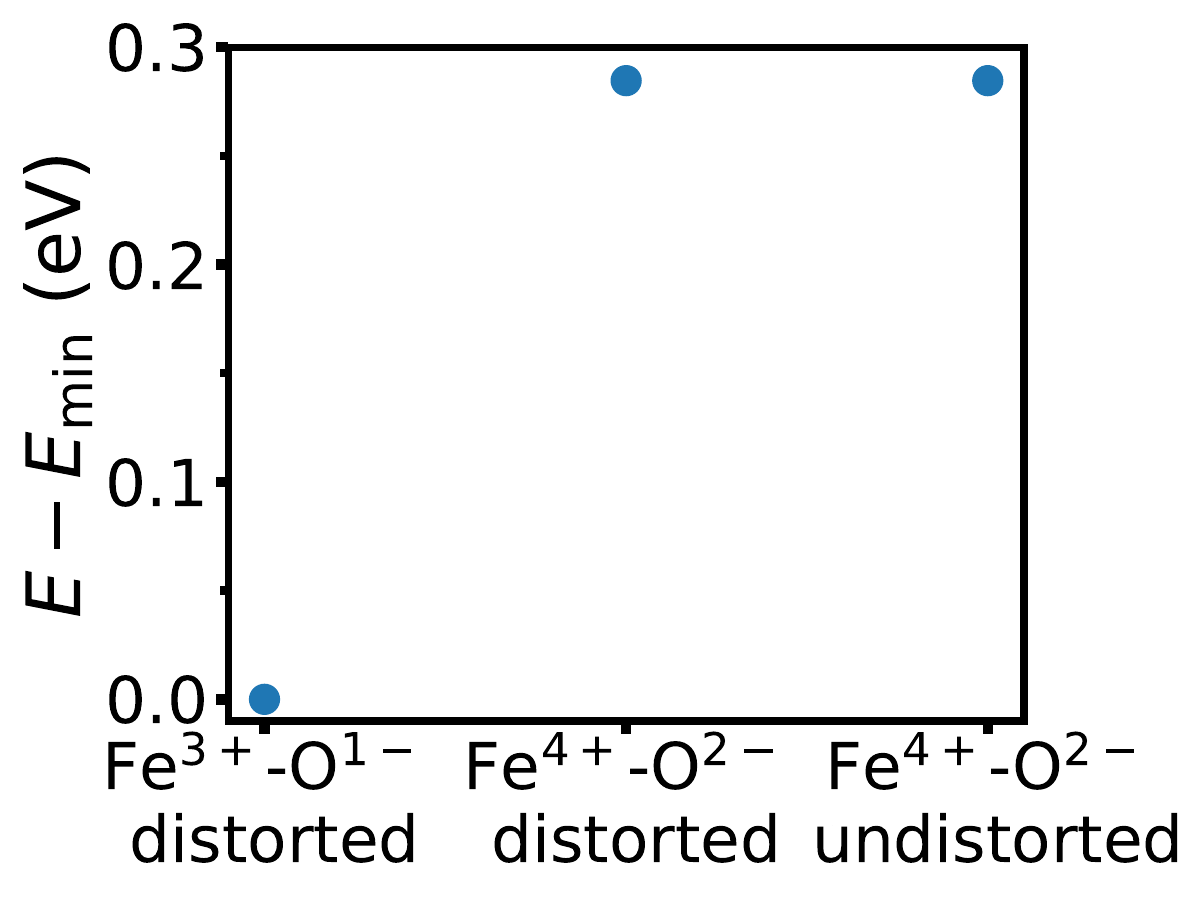}\label{fig:2-1}}
  \hfill
  \subfigure[]{
    \includegraphics[width=0.48\columnwidth]{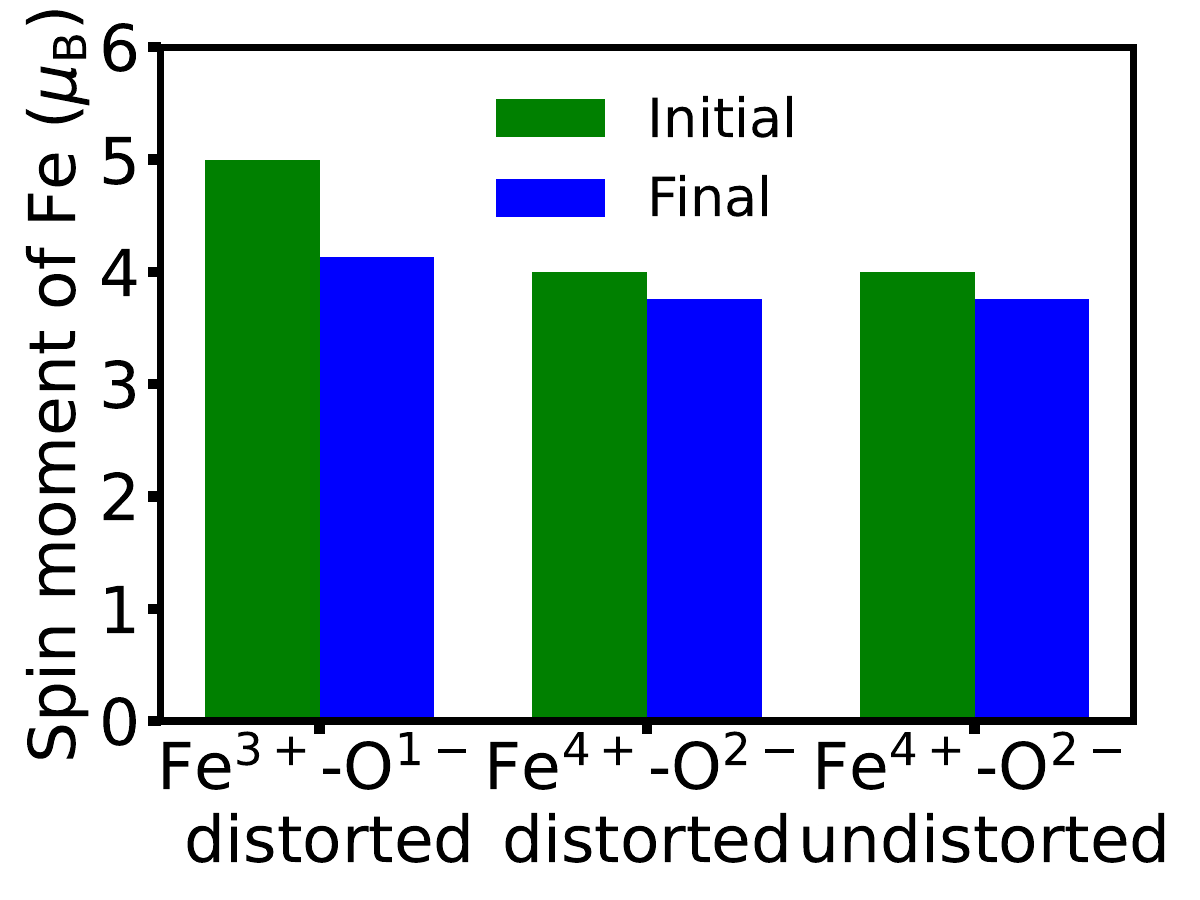}\label{fig:2-2}}
  \caption{(Color online) 
(a) Total energies for the Fe\(^{3+}\)--O\(^{-}\) in the
polaron-distorted lattice (``Fe\(^{3+}\)--O\(^{-}\) distorted''), the high-spin Fe\(^{4+}\) state constrained
to the same distorted lattice (``Fe\(^{4+}\)--O\(^{2-}\) distorted''), and Fe\(^{4+}\) in the undistorted pristine
lattice (``Fe\(^{4+}\)--O\(^{2-}\) undistorted''). The Fe\(^{3+}\) + O\(^{-}\) configuration is lower in energy by
\( 0.28\) (eV) than either Fe\(^{4+}\)--O\(^{2-}\) configuration.
(b) Local spin moments on the Fe atom for the
same three cases, comparing the values corresponding to the imposed initial occupations (green bars)
and the converged, fully relaxed electronic structures (blue bars).
}
  \label{fig:fig-3}
\end{figure}

\textit{Oxidation state of Fe--}
We next determine how an oxidizing hole is accommodated when Fe substitutes Ti in BaTiO$_3$. 
Starting from the relaxed O-hole-polaron geometry, the Ti atom adjacent to the polaron-hosting oxygen is replaced by Fe. Two electronic configurations are then initialized. For Fe$^{3+}$--O$^{-}$, we impose a high-spin Fe $d^5$ state and localize the compensating hole on the neighboring O-2$p$ manifold. For formal Fe$^{4+}$, we impose a high-spin Fe $d^4$ state without an O-centered hole. In both cases, the occupation constraints are used only to access the desired local minimum; they are subsequently released and the electronic structure is relaxed again. The resulting solutions therefore correspond to stationary states of the unconstrained DFT+$U$ functional rather than artifacts of the initialization procedure. 

Figure~\ref{fig:fig-3}(a) compares the total energies. The \mbox{Fe$^{3+}$--O$^{-}$} configuration is lower in energy than the formal Fe$^{4+}$ state by $0.28$ eV. Repeating the Fe$^{4+}$ calculation in the undistorted rhombohedral BaTiO$_3$ lattice gives essentially the same energy as in the polaron-distorted lattice, showing that the stabilization is not merely a consequence of the initial distortion. Instead, the energy gain reflects the intrinsic preference to accommodate the oxidizing hole on oxygen rather than on Fe.

The local magnetic moments, shown in Fig.~\ref{fig:fig-3}(b), support this assignment. After removal of the occupation constraints, the Fe moment in the Fe$^{3+}$--O$^{-}$ state relaxes to $4.13~\mu_{\mathrm{B}}$, consistent with  a
high-spin Fe$^{3+}$ center with substantial Fe--O covalency \cite{Zorko2015,IVANOV20123253,boateng2017dft}. The polaron-hosting oxygen carries a large moment of about $0.75~\mu_{\mathrm{B}}$, and its projected density of states exhibits a spin-polarized O-2$p$ in-gap level, as shown in Fig.~S2(a). By contrast, the two formal Fe$^{4+}$ configurations have smaller Fe moments of about $3.76~\mu_{\mathrm{B}}$, weak spin polarization on neighboring oxygen atoms, and no distinct O-centered in-gap state [Figs.~S2(b) and S2(c)]. 
The hole is therefore more delocalized over the Fe--O network in the formal Fe$^{4+}$ solutions.

To test whether the Fe$^{3+}$--O$^{-}$ complex is a stable bound state, we evaluate its association energy relative to separated Fe$_{\mathrm{Ti}}^{'}$ and an isolated O-centered hole polaron using Eq.~(S2). The resulting binding energy is
\begin{equation}
E_{\mathrm{bind}}=-0.17~\mathrm{eV},
\end{equation}
showing that association between Fe$^{3+}$ and the oxygen hole polaron is
energetically favorable. This result is insensitive to the precise value of the
Fe Hubbard parameter: varying $U_{\mathrm{Fe}\,3d}$ from $4.5$ to $5.5$ eV gives
$E_{\mathrm{bind}}=-0.17$~eV and $-0.16$~eV, respectively
\cite{Moore2024}. A larger $4\times4\times4$ supercell yields
$E_{\mathrm{bind}}=-0.16$ eV, confirming that residual finite-size effects are
small.

These results show that oxidation of Fe-doped BaTiO$_3$ favors a ligand-hole
configuration over a formal Fe$^{4+}$ center. The relevant local
charge-compensation mechanism is therefore the formation of a bound
Fe$^{3+}$--O$^{-}$ hole-polaron complex, rather than oxidation of Fe alone.

\textit{EPR Spectroscopy Aspect--}
Electron-paramagnetic resonance (EPR) provides a sensitive probe of
paramagnetic Fe-related centers in BaTiO$_3$, but it does not offer a direct
one-to-one test for every possible Fe oxidation state. In particular, high-spin
Fe$^{3+}$ is readily detected, whereas Fe$^{4+}$ may be EPR silent or difficult
to identify unambiguously, depending on its spin state, covalency, relaxation
rates, and line broadening. Consequently, the absence of a resolved Fe$^{4+}$
signal cannot, by itself, prove that Fe$^{4+}$ is absent.

Nevertheless, the available EPR literature provides a clear experimental
constraint: the dominant Fe-related signals in BaTiO$_3$ and related titanate
perovskites are consistently assigned to Fe$^{3+}$-based centers, often in
distorted octahedral environments and frequently associated with oxygen-related
defects~\cite{schwartz1993electron,possenriede1992paramagnetic,laguta2005electron, theerthan2012magnetic}. These
observations indicate that, under a broad range of preparation and measurement
conditions, the experimentally accessible Fe population is predominantly
trivalent rather than clearly high-valent.

This spectroscopic picture is consistent with the present calculations. We find
that the formal Fe$^{4+}$ configuration is not the lowest-energy oxidized state.
Instead, the oxidizing hole preferentially localizes on a neighboring oxygen
ligand, forming a Fe$^{3+}$--O$^{-}$ bound-polaron complex. In this configuration
the Fe site retains a high-spin Fe$^{3+}$-like moment, while the neighboring
oxygen carries the ligand-hole spin density. Thus, the calculated ground state
naturally explains why EPR experiments predominantly observe Fe$^{3+}$-based
centers even when oxidizing conditions would nominally favor Fe oxidation.

The role of EPR in the present work is therefore not to exclude Fe$^{4+}$ by
non-observation, but to provide an experimental benchmark for the dominant
detectable Fe-related defect chemistry. Our results offer a microscopic
interpretation of this benchmark: charge compensation in oxidized Fe-doped
BaTiO$_3$ is more favorably achieved by ligand-hole formation on oxygen than by
formation of a stable bulk Fe$^{4+}$ center.
Further aspects of EPR data (Mn$^{3+}$--O$^{-}$ ligand-hole models), co-doping effects (Mn--Nb, Pr--Mn), hexagonal polymorphs, charge-transition-level comparisons across perovskites, and microstructural aspects are discussed in detail in the Supplemental Material
\cite{bottcher2005evaluation,muller1959electron,scharfschwerdt1999fermi,mukherjee2021structure,lu2011tetragonal,Klein2023Fermi}.

\textit{Conclusion--}
We have shown that oxidation in Fe-doped BaTiO$_3$ is not most favorably
accommodated by forming a formal Fe$^{4+}$ center, but by transferring the
oxidizing hole to the oxygen sublattice. Using DFT+$U$ with occupation-matrix
control, we first established that rhombohedral BaTiO$_3$ can stabilize an
O-centered small hole polaron. The localized hole forms a spin-polarized O-2$p$
state in the band gap, located about $1.0$ eV above the valence-band maximum,
and is energetically favored over the delocalized valence-band-hole solution.

When Fe substitutes Ti, this oxygen hole becomes bound to the Fe impurity. The
resulting Fe$^{3+}$--O$^{-}$ complex is lower in energy than the formal
Fe$^{4+}$ configuration by $0.28$ eV. Its negative binding energy,
$E_{\mathrm{bind}}=-0.17$~eV, demonstrates that the complex is a stable
defect-associated state rather than a consequence of the occupation constraint or
the initial polaronic distortion. The stability of this configuration is robust
with respect to the Fe Hubbard parameter and supercell size.

This result changes the interpretation of oxidation in Fe-doped
BaTiO$_3$. In a purely ionic picture, lowering the Fermi level would be expected
to oxidize Fe$^{3+}$ to Fe$^{4+}$. Our calculations instead show that the hole is
preferentially accommodated as a ligand hole on oxygen, while Fe remains largely
trivalent. The relevant oxidized defect is therefore better described as a
Fe$^{3+}$--O$^{-}$ bound-polaron complex, or equivalently as a ligand-hole
configuration, rather than as an isolated high-valent Fe center.

This picture is consistent with the spectroscopic phenomenology of Fe-doped
BaTiO$_3$. EPR experiments predominantly detect Fe$^{3+}$-based centers, often
associated with oxygen-related defects, while Fe$^{4+}$ is not available as a
straightforward EPR benchmark. The present results provide an
explanation for this observation: the lowest-energy oxidized configuration
retains a Fe$^{3+}$-like center and places the compensating spin and charge on a
neighboring oxygen ligand.

Thus, our findings identify oxygen hole polarons as active
charge-compensation centers in acceptor-doped ferroelectric perovskites. Such
polaronic defects can limit Fermi-level shifts, modify carrier concentrations,
and influence electronic conductivity under oxidizing conditions. The mechanism
demonstrated here for Fe-doped BaTiO$_3$ is therefore likely relevant beyond this
specific material, particularly in transition-metal-doped oxides where formal
high-valent cation states compete with ligand-hole formation. Oxygen polarons
should thus be treated not as secondary electronic excitations, but as relevant
defect species in the context of Fermi-level engineering of oxide electroceramics.

\textit{Acknowledgement--}
Authors acknowledge financial support from the Collaborative Research Center FLAIR (Fermi level engineering applied to oxide electroceramics), funded by the German Research Foundation (DFG) under Project-ID No. 463184206–SFB 1548.
Lichtenberg and Paderborn Supercomputing centers are gratefully acknowledged by MA. MA sincerely thanks Vasilios Karanikolas for his discussions and suggestions.
\bibliographystyle{apsrev4-1}
\bibliography{bib.bib} 
\end{document}